\documentclass[%
 reprint,
nofootinbib,
  showpacs,
 amsmath,amssymb,
 aps,
prb,
longbibliography,
]{revtex4-2}

\usepackage[dvipdfmx]{graphicx}
\usepackage{bmpsize}
\usepackage{dcolumn}% Align table columns on decimal point
\usepackage{bm}% bold math
\usepackage{multirow}
\usepackage{longtable}
\usepackage[utf8]{inputenc}   % Correct display of \UTF{00F6},\UTF{00FC},\UTF{00E4} and all other characters
\usepackage[T1]{fontenc}      % Important for anything that uses more than ASCII
\usepackage[english]{babel}   % Language ! e.g. ngerman, english
\usepackage{lmodern}          % Beautiful fonts especially for pdfs
\usepackage{csquotes}         % Beautiful quotes with \enquote{Text}
\usepackage{amsmath}          % Better and more beautiful mathematical formulas
\usepackage{mathtools}        % Fix and expand amsmath
\usepackage{amstext}          % \text{} macro in math mode
\usepackage{amssymb}          % Special mathematical symbols
\usepackage{subdepth}         % Unify maths subscript height
\usepackage{xcolor}           % All the colors
\usepackage[
  separate-uncertainty=true,
  per-mode=reciprocal,
  binary-units=true,
  range-phrase=--]{siunitx}   % Beautiful SI units
\usepackage{braket}           % For quantum mechanical Bra-Kets
\usepackage{graphicx}         % Important for adding graphics
\usepackage[pass]{geometry}   % For changing page layout and geometry
\usepackage[capitalize]{cleveref}

\usepackage{braket}
\usepackage{color}
\usepackage{ulem}

\usepackage{times}
\usepackage{booktabs}
\begin{document}

\preprint{APS/123-QED}

\title{First-principles investigation of magnetic and transport properties in hole-doped shandite compounds Co$_3$In$_x$Sn$_{2-x}$S$_2$}% Force line breaks with \\

\author{Yuki Yanagi$^1$, Junya Ikeda$^2$, Kohei Fujiwara$^2$, Kentaro Nomura$^{2,3}$, Atsushi Tsukazaki$^{2,3,4}$ and Michi-To Suzuki$^{1,5}$}
\affiliation{
$^1$Center for Computational Materials Science, Institute for Materials Research, Tohoku University, Sendai, Miyagi, 950-8577, Japan \\
$^2$Institute for Materials Research, Tohoku University, Sendai Miyagi 950-8577, Japan \\
$^3$Center for Spintronics Research Network, Tohoku University, Sendai, Miyagi 980-8577, Japan \\
$^4$Center for Science and Innovation in Spintronics (CSIS), Core Research Cluster, Tohoku University, Sendai, Miyagi 980-8577, Japan \\
$^5$Center for Spintronics Research Network, Graduate School of Engineering Science, Osaka University, Toyonaka, Osaka 560-8531, Japan
}
%\affiliation{%
%\affiliation{
%\affiliation{
%\affiliation{%

\begin{abstract}
  Co-based shandite Co$_3$Sn$_2$S$_2$ is a representative example of magnetic Weyl semimetals showing rich transport phenomena. 
   We thoroughly investigate  magnetic 
   and transport properties of hole-doped shandites Co$_3$In$_x$Sn$_{2-x}$S$_2$  by    first-principles calculations. 
The calculations reproduce
    nonlinear  reduction of anomalous Hall conductivity 
   with doping In for Co$_3$Sn$_2$S$_2$, as     reported in experiments, against the linearly decreased ferromagnetic moment within virtual crystal approximation. 
We show that a drastic change in the band parity character of Fermi surfaces, attributed to the nodal rings lifted energetically with In-doping, leads to strong enhancement of anomalous Nernst conductivity with reversing its sign in Co$_3$In$_x$Sn$_{2-x}$S$_2$.

\end{abstract}

\maketitle

%\tableofcontents

\section{Introduction}

 Novel charge and spin transports have attracted growing interests in the contexts of topological phases of matter.
The anomalous Hall effect (AHE) and anomalous Nernst effect (ANE) 
 are classical examples of such phenomena and have been studied for many years~\cite{Nagaosa_RevModPhys.82.1539}. 
 The AHE (ANE) is characterized by a transverse charge current flow $\bm{j}$ induced by an applied electric field $\bm{E}$ (thermal gradient $\bm{\nabla}T$) in the absence of an external magnetic field  as follows, 
\begin{align}
j_a=\sum_b  \left[\sigma^{\mathrm{A}}_{ab}E_b+\alpha^{\mathrm{A}}_{ab}\left(-\nabla_b T\right)  \right], 
\end{align}
where antisymmetric tensors $\sigma^{\mathrm{A}}_{ab}$ and $\alpha^{\mathrm{A}}_{ab}$ are  anomalous Hall conductivity (AHC) and anomalous Nernst conductivity (ANC), respectively,   with $a,b=x,y,z$ . 
 Since the AHC and ANC are closely related to  topological properties of  electronic structures via Berry curvature~\cite{Nagaosa_RevModPhys.82.1539,Xiao_RevModPhys.82.1959,Armitage_RevModPhys.90.015001,Tokura_NatRevPhys_2019,Felser_ANCreview2020}, 
   significant effort has been devoted to exploring such anomalous transports in topological quantum matters with time-reversal symmetry breaking.  
  Recently, large anomalous Hall and Nernst responses have been observed in magnetic Weyl semimetals including Co$_3$Sn$_2$S$_2$ and Co$_2$MnGa~\cite{Liu_shandite_2018,Wang_shandite_2018,Guin_AdvMat.31.1806622,Sakai_NatPhys2018Co2MnGa,Guin_NPG2019}.

 Shandite compound Co$_3$Sn$_2$S$_2$ is a half-metallic ferromagnet with transition temperature  $T_{\mathrm{c}}\sim \SI{177}{\kelvin}$ and saturated moment $M\sim 0.3\,\mu_{\mathrm{B}}$ per Co atom~\cite{Weihrich_ZAnorAllgChem2016,Holder_PhysRevB.79.205116,Schnelle_PhysRevB.88.144404,Fujiwara_2019_JJAP}. 
This material is a  representative example of magnetic Weyl semimetals since, according to the angle resolved photoemission spectroscopy (ARPES)  and electronic structure calculations, Weyl nodes are located near the Fermi energy~\cite{Liu_shandite_2018,Wang_shandite_2018,Liu_Science1282_ARPES_Shandite_2019}. 
    In  Weyl semimetals, the divergent behavior of the Berry curvature at the Weyl nodes can give rise to characteristic physics, e.g., the chiral magnetic effect and the emergence of the anomalous surface states called Fermi arcs~\cite{Armitage_RevModPhys.90.015001}.
    The existence of the anomalous surface states in Co$_3$Sn$_2$S$_2$ has been confirmed by the ARPES and scanning tunneling spectroscopy (STS) measurements~\cite{Liu_Science1282_ARPES_Shandite_2019,Jiao_PhysRevB.99.245158,Morali_Science.365.1286}, 
     supporting the existence of the Weyl nodes near the Fermi energy.  
     
  Co$_3$Sn$_2$S$_2$ is expected to be a potential candidate for the thermoelectric devise applications due to 
 the characteristic  transport phenomena with the large AHC and ANC, reaching $\sigma^{\mathrm{A}}_{xy}\sim 500$-$\SI{1130}{\siemens \per \centi \meter}$~\cite{Liu_shandite_2018,Wang_shandite_2018} and $\alpha^{\mathrm{A}}_{xy}\sim 2$-$\SI{10}{\ampere\per\kelvin\per\metre}$~\cite{Guin_AdvMat.31.1806622,Yang_PhysRevMaterials.4.024202}, respectively.
 Previous theoretical studies imply the close relation between  these anomalous transports and topological bands such as Weyl nodes and nodal rings~\cite{Nagaosa_RevModPhys.82.1539,Xiao_RevModPhys.82.1959,Armitage_RevModPhys.90.015001,Felser_ANCreview2020,Liu_shandite_2018,Wang_shandite_2018,Burkov_PhysRevLett.113.187202,Ghimire_PhysRevResearch.1.032044,Minami_PhysRevB.102.205128}.
 Controlling pressure, temperature and chemical composition often affect 
   transport properties. It is known that, in Co$_3$Sn$_2$S$_2$, applying pressure suppresses the AHE, 
 leading to the AHC and Hall angle comparable to those in   conventional ferromagnetic metals~\cite{Chen_PhysRevB.100.165145,Liu_PhysRevMaterials.4.044203,Guguchia_NatCommun2020}.

 It has also been shown experimentally that  the chemical substitution have a considerable impact on  the magnetic and transport properties. 
 So far, the effects of substitution of Fe and Ni for Co and of In for Sn  have been  investigated~\cite{Weihrich_ZAnorAllgChem2016,Kassem_JPSJ2016,Corps_shandite_2015,Jianlei_adfm.202000830,Thakur_ChemMater.32.1612,Zhou_PhysRevB.101.125121}, 
 where Ni substitution corresponds to the electron doping, while Fe and In substitutions correspond to the hole doping.
  Irrespective of doped elements, similar doping dependence of magnetic properties has been confirmed, that is, 
   the monotonic decrease of the transition temperature and magnetic moment with increasing doping content.  
  The doping effects on transport properties, on the other hand,  show more complicated behaviors. 
  In the case of the Ni substitution, the AHC decreases monotonically with the doping content~\cite{Thakur_ChemMater.32.1612}, while in the cases of  Fe and In substitutions, 
  AHC is enhanced  for relatively small doping and decreases for large doping~\cite{Jianlei_adfm.202000830,Zhou_PhysRevB.101.125121}.  
  From the theoretical side,  the effects of substitution atoms on magnetic properties and/or electric transport properties 
  have been previously investigated based on the first-principles calculations with the supercell approach for specific doping concentrations~\cite{Corps_shandite_2015,Rothballer_C4RA03800B,Jianlei_adfm.202000830,Thakur_ChemMater.32.1612}. 
  A systematic investigation of the atom substitution effects on the thermoelectric transport, i.e., Nernst effect has not been performed. 
 Here, we present a systematic investigation of the effects of In-doping into Sn sites on the  thermoelectric transports as well as magnetic properties based on the first-principles method.

\section{Methods}

 We perform the density functional calculation for Co$_3$In$_x$Sn$_{2-x}$S$_2$ by using the \textsc{wien2k} code~\cite{wien2k_2001,wien2k_2020}. 
  Sn/In substitution effects  are treated within the virtual crystal approximation (VCA) and with the lattice parameters adopted from the experimental values of Co$_3$Sn$_2$S$_2$ in ref.~\cite{Corps_shandite_2015}, independently of In content $x$.
  Sn atoms in Co$_3$Sn$_{2}$S$_{2}$ occupy two inequivalent Wyckoff positions, intra and inter Co Kagom\'{e} layer sites, and in the present study, Sn atoms at the inter-layer sites are substituted by In atoms for the proper description of the paramagnetic insulating phase at $x=1.0$~\cite{Rothballer_C4RA03800B}. 
  We generate the maximally localized Wannier function to construct an effective tight-binding Hamiltonian from the obtained electronic structures  with  \textsc{wannier90} package~\cite{Marzari_RevModPhys.84.1419,Pizzi_2020} through the \textsc{wien2wannier} interface~\cite{Kunes_20101888}. 
 We explicitly include  Co-$3d$, Sn/In-$5s$, $5p$, and S-$3p$ orbitals for the Wannier model.  
 Based on the Wannier models, we  investigate the intrinsic contributions to electric and thermoelectric transports, which are determined from purely electronic band structures, in Co$_3$In$_x$Sn$_{2-x}$S$_2$ 
 by calculating the AHC and ANC with use of the following Kubo formulae~\cite{Nagaosa_RevModPhys.82.1539,Haldane_PhysRevLett.93.206602,Wang_PhysRevB.74.195118,Xiao_PhysRevLett.97.026603},
\begin{align}
&\sigma^{\mathrm{A}}_{ab} \left(\mu,T \right)= -\frac{e^2}{\hbar}\int_{\mathrm{BZ}} \!\! \frac{d\bm{k}}{\left(2\pi \right)^3}  \sum_n f\left(\varepsilon_{n\bm{k}}\right) \Omega_{n,ab} \left(\bm{k}\right), \label{Eq:AHC} \\
& \alpha^{\mathrm{A}}_{ab} \left(\mu,T \right)= \frac{ek_{\mathrm{B}}}{\hbar}\int _{\mathrm{BZ}} \!\! \frac{d\bm{k}}{\left(2\pi \right)^3}  \sum_n s\left(\varepsilon_{n\bm{k}}\right)\Omega_{n,ab} \left(\bm{k}\right), \label{Eq:ANC}
\end{align}
where $e$, $k_\mathrm{B}$, $\hbar$ and $\varepsilon_{n\bm{k}}$ are the positive elementary charge, Boltzmann constant, reduced Planck constant and one-particle energy with band index $n$ and wave vector $\bm{k}$, respectively.  
The Fermi distribution function $f(\varepsilon)$ and the entropy density $s(\varepsilon)$ are given as  $f(\varepsilon)=(e^{\frac{\varepsilon-\mu}{k_\mathrm{B}T}}+1)^{-1}$ and 
$s(\varepsilon)= -f(\varepsilon)\log f(\varepsilon)- \left[1-f(\varepsilon)\right]\log \left[1-f(\varepsilon)\right]  $.  
 $\Omega_{n,ab}(\bm{k})$ is the Berry curvature for band $n$, which is expressed as follows: 
\begin{align}
\Omega_{n,ab}(\bm{k})=-2\hbar^2\mathrm{Im}\sum_{m(\ne n)}\frac{\Braket{n\bm{k}|v_{a}|m\bm{k}}\Braket{m\bm{k}|v_{b}|n\bm{k}}}{\left(\varepsilon_{n\bm{k}}-\varepsilon_{m\bm{k}}\right)^2}, 
\end{align}
where $v_{a}$ is the velocity operator along $a$-direction, and $\Ket{n\bm{k}}$ is the Bloch state with band index $n$ and wave vector $\bm{k}$. 
In the actual numerical calculation, we perform the $\bm{k}$-integration in Eqs.~(\ref{Eq:AHC}) and (\ref{Eq:ANC}) as the discrete $\bm{k}$-summation on $250^3$ grids in the first Brillouin zone.

\begin{figure}[tb!]
\begin{center}
\includegraphics[width=1.0 \hsize]{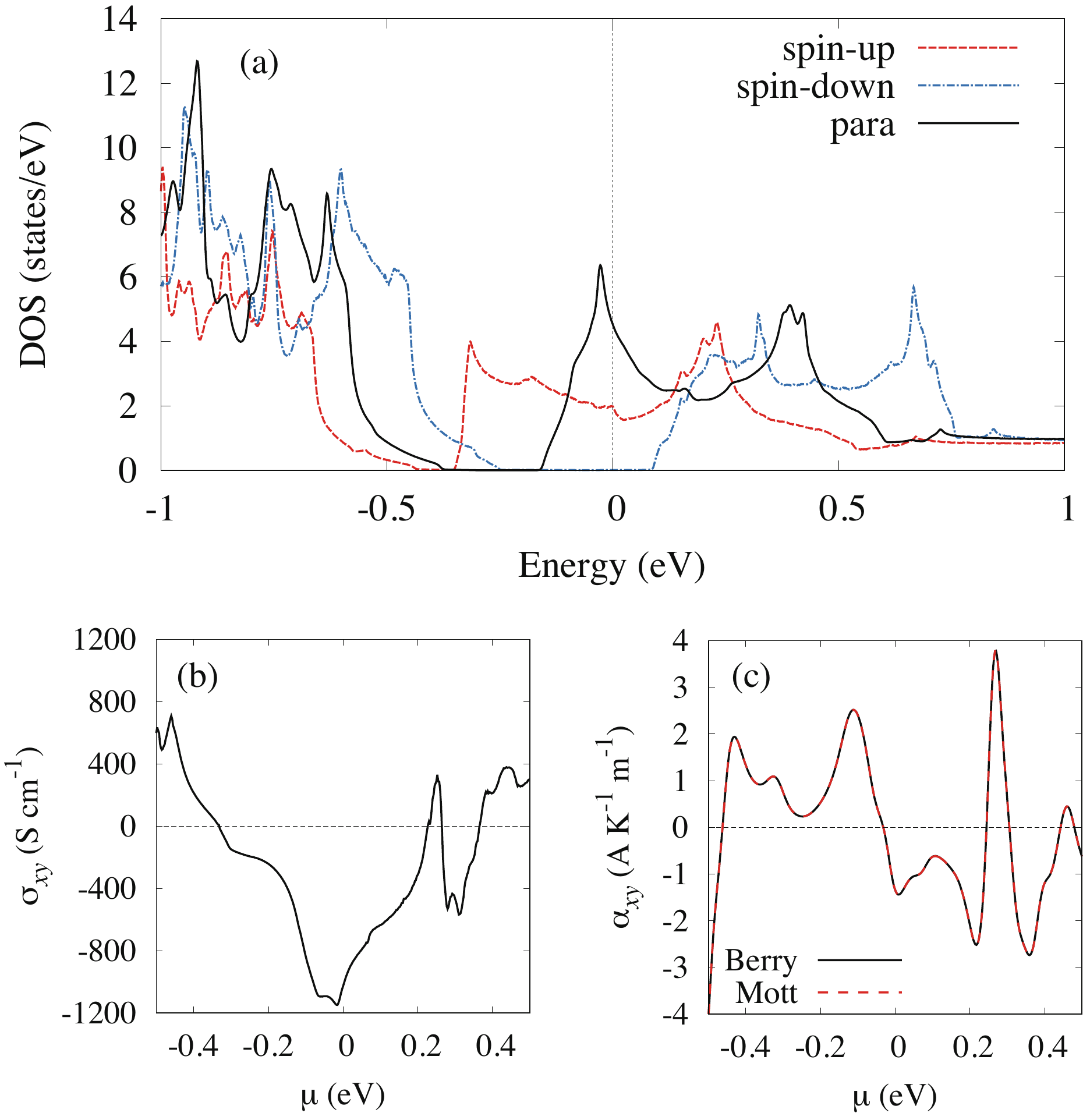}
\caption{
\label{Fig:dos}
(a) DOS for Co$_3$Sn$_2$S$_2$ in paramagnetic and ferromagnetic states. The solid and dashed (dot-dashed) lines represent DOS per spin in the paramagnetic state and  that with spin-up (down) in the ferromagnetic state, respectively.
The chemical potential dependences of (b) AHC at $k_{\mathrm{B}}T=0$ and (c) ANC at $k_{\mathrm{B}}T=\SI{0.01}{\electronvolt}$.  
 In panel (c), the solid and dashed lines represent the ANCs calculated via Berry phase formula in Eq.~(\ref{Eq:ANC}) and generalized Mott formula in Eq.~(\ref{Eq:Mott1}), respectively.  
}
\end{center}
\end{figure}

\begin{figure}[tb]
\begin{center}
\includegraphics[width=1.0 \hsize]{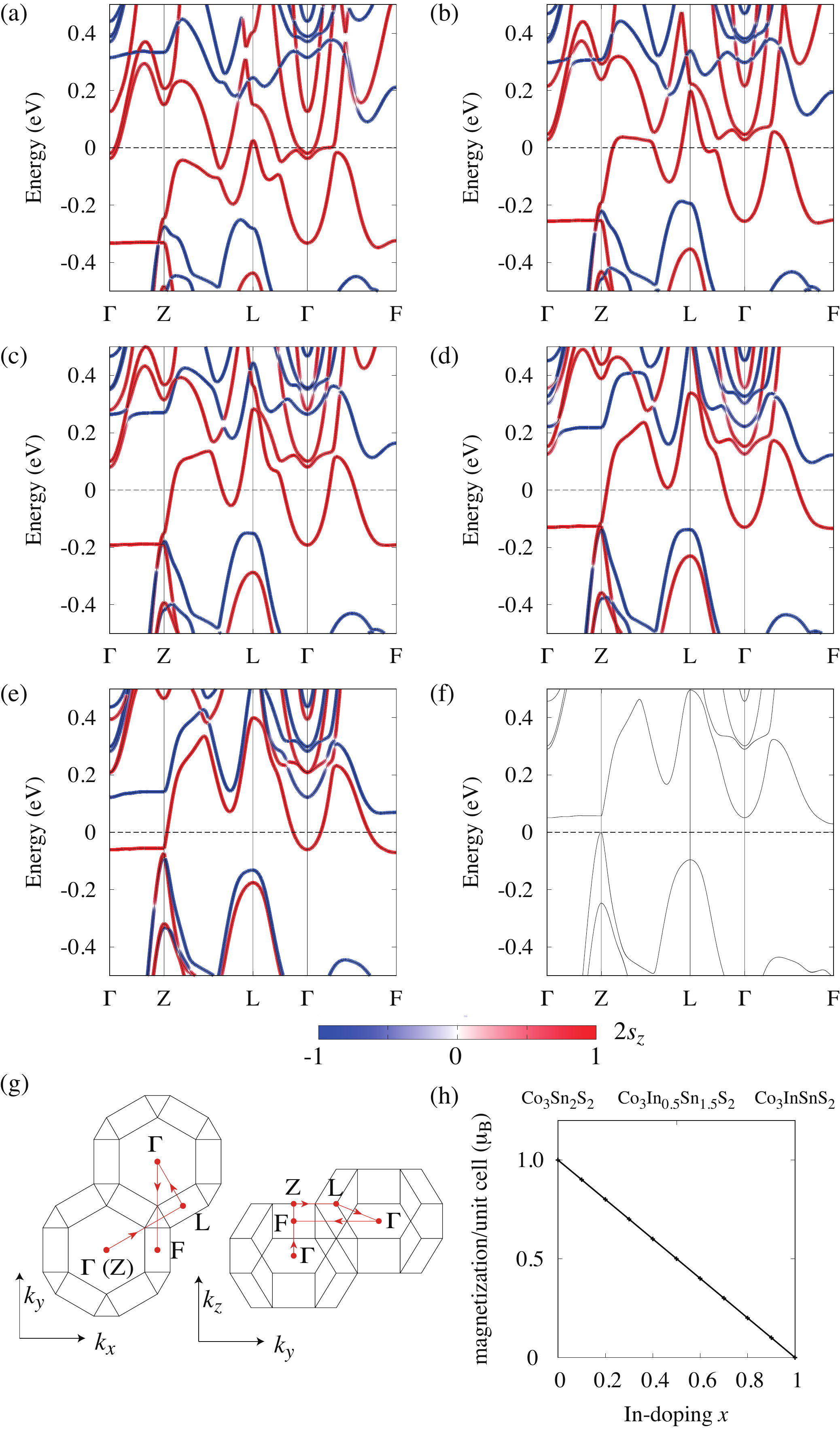} 
\caption{
\label{Fig:band}
 (a)-(f) Band structures for Co$_3$In$_x$Sn$_{2-x}$S$_2$ along high symmetry $\bm{k}$-lines, where the color map represents the spin density along $z$-axis. 
 (g) $\bm{k}$-path on which the band structures are plotted in panels (a)-(f). (h) In-doping dependence of the net magnetization. 
}
\end{center}
\end{figure}

\begin{figure}[t]
\begin{center}
\includegraphics[width=1.0 \hsize]{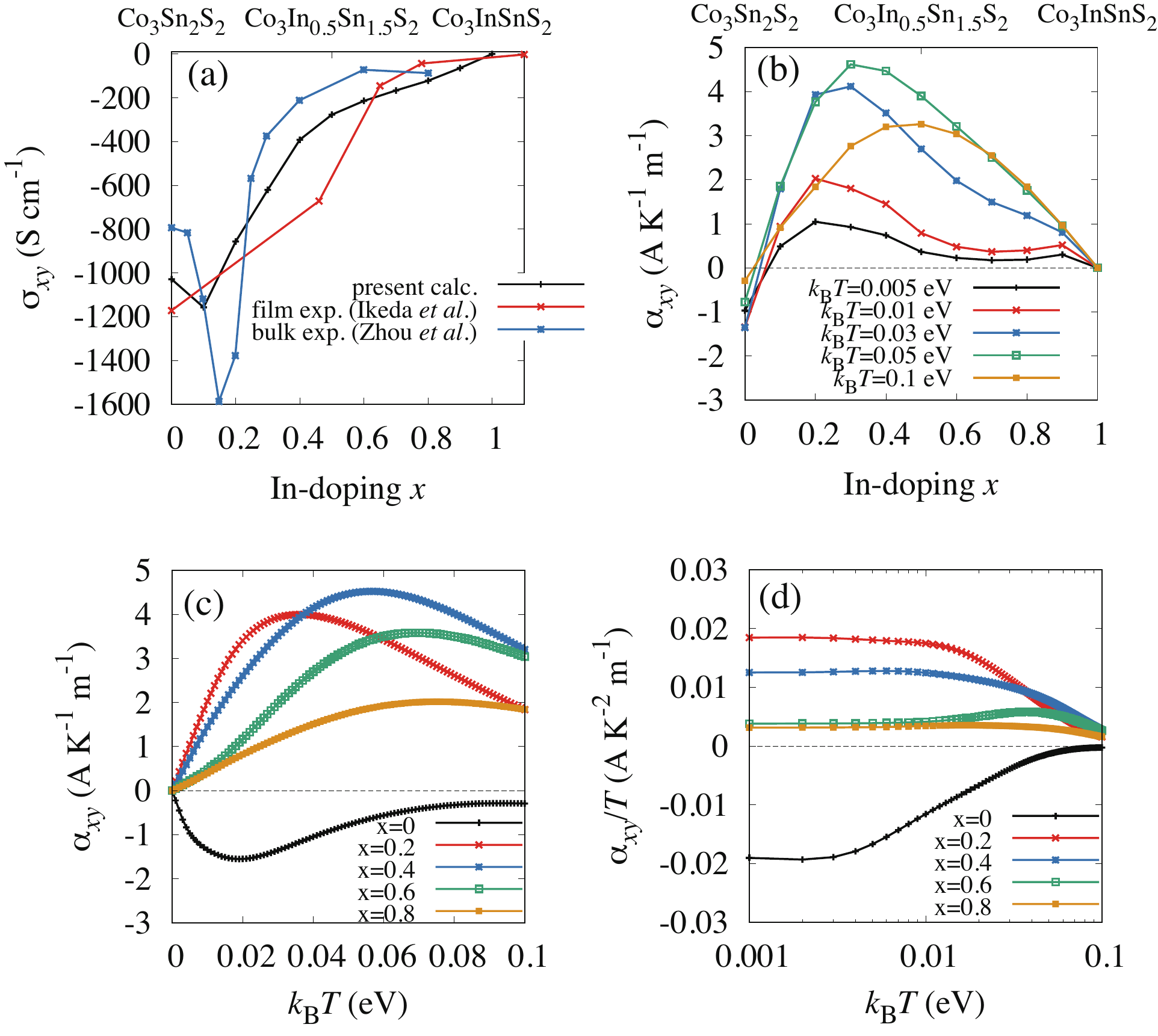}
\caption{
\label{Fig:AHE}
(a) Doping dependence of AHC $\sigma^{\mathrm{A}}_{xy}$ at $k_{\mathrm{B}}T=0$ and (b) that of ANC  $\alpha^{\mathrm{A}}_{xy}$ at finite temperatures. 
 (c) Temperature dependence of   $\alpha^{\mathrm{A}}_{xy}$ and (d) that of  $\alpha^{\mathrm{A}}_{xy}/T$ at various In-doping concentrations.
In panel~(a),  experimental data extracted from refs.~\cite{Zhou_PhysRevB.101.125121,Ikeda_AHE} are plotted together with calculated results for comparison.
 }
\end{center}
\end{figure}

\begin{figure*}[t]
\begin{center}
\includegraphics[width=1.0 \hsize]{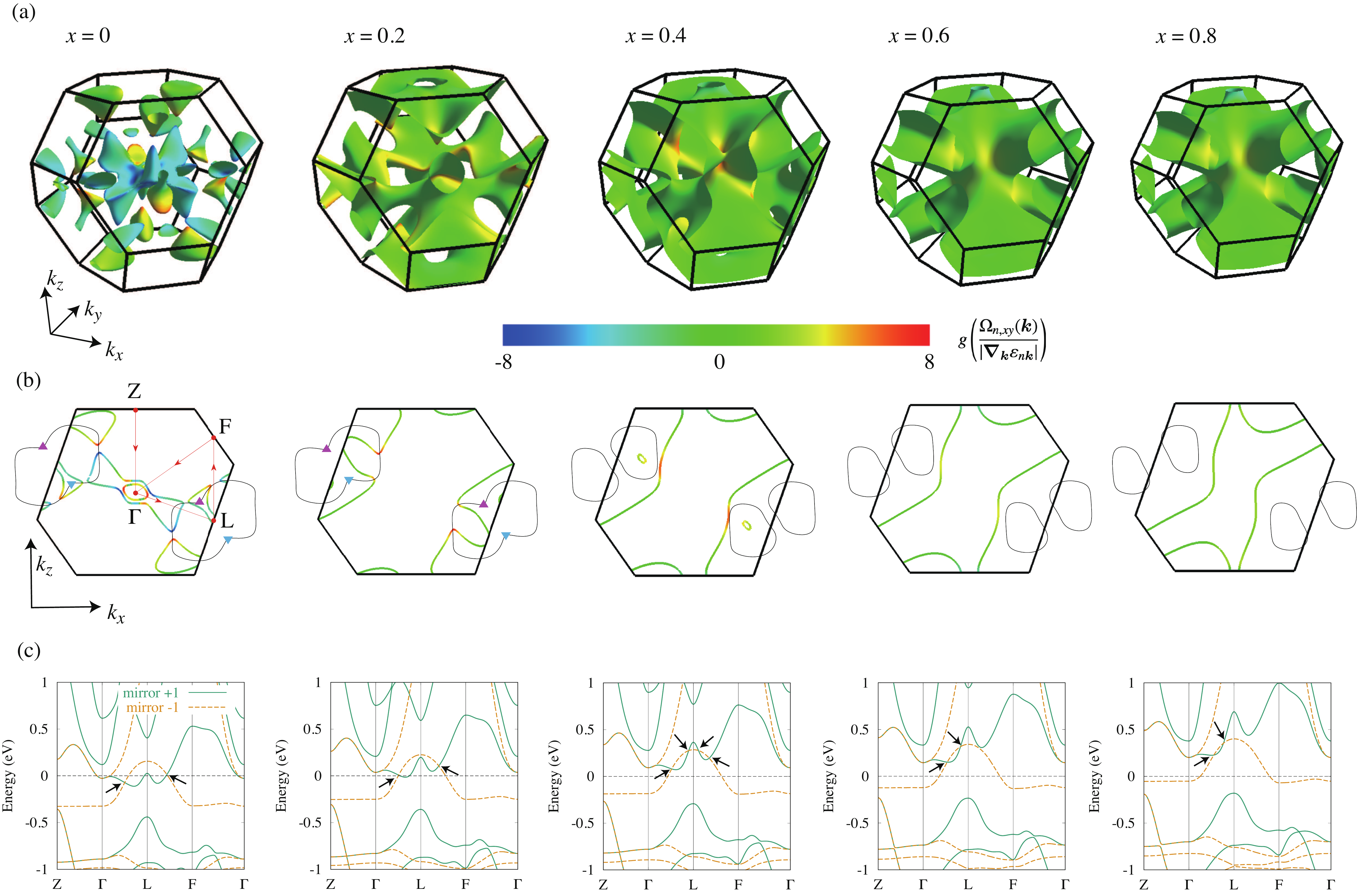}
\caption{
\label{Fig:FS_Berry}
(a) Fermi surfaces for Co$_3$In$_x$Sn$_{2-x}$S$_2$, where the color map represents the intensity of $\frac{\Omega_{n,xy}(\bm{k})}{|\bm{\nabla}_{\bm{k}}\varepsilon_{n\bm{k}}|}$ 
through the logarithmic function $g(x)\equiv \mathrm{sgn}(x)\log (1+|x|)$. 
(b) Those on $k_x$-$k_z$ plane at $k_y=0$. The black solid line represents the nodal lines in the absence of  spin-orbit coupling. The upper (lower) triangles on the nodal lines denote the Weyl nodes with topological charge $+1$ ($-1$) in the presence of spin-orbit coupling. The $\bm{k}$-path along which band structures plotted in panel (c) is shown by red arrows in the left most panel.
(c) Majority spin band structures in the absence of the spin-orbit coupling. The solid and dashed lines represent the energy bands with eigen values $+1$ and $-1$  of mirror symmetry operator on $k_x$-$k_z$ plane, respectively. Band crossings corresponding to nodal lines are shown by arrows.
Figures~(a) and (b) are created with use of \textsc{fermisurfer} code~\cite{KAWAMURA2019197}.
 }
\end{center}
\end{figure*}

\section{Results}
 First, we show the results for $x=0$, i.e., Co$_3$Sn$_2$S$_2$.  
  Figure~\ref{Fig:dos}(a) shows the density of states (DOS) for Co$_3$Sn$_2$S$_2$ both in the paramagnetic and ferromagnetic states. 
   The DOS in the paramagnetic state has a gap between  $\SI{-0.31}{\electronvolt}$ to $\SI{-0.17}{\electronvolt}$ and 
    the states from  $\SI{-0.17}{\electronvolt}$ to the Fermi energy $E_{\mathrm{F}}$  are occupied by one electron per unit cell, resulting in the fully-polarized magnetic moment $\sim 1\mu_{\mathrm{B}}$ per unit cell in the ferromagnetic calculation.  
The exchange splitting pushes down the up-spin DOS, leading to the Fermi energy around the dip of the up-spin DOS, and pushes up the down-spin DOS, resulting the Fermi energy in a gap of down-spin DOS. As a consequence, the half-metallic state is realized. 

 The electric and thermoelectric transport properties in Co$_3$Sn$_2$S$_2$ are now discussed.  Since the Berry curvature and  ferromagnetic moment  have the same symmetry properties under the magnetic point group,  
    the AHC and ANC can be finite in the ferromagnetic state~\cite{Wang_PhysRevB.74.195118,Birss_symmetry,Seeman_PhysRevB.92.155138,Watanabe_PhysRevB.98.245129,Hayami_PhysRevB.98.165110,Suzuki_PhysRevB.95.094406}. 
     In the present case, since the magnetic point group of the ferromagnetic phase is $\bar{3}m'$, 
     only $\sigma^{\mathrm{A}}_{xy}=-\sigma^{\mathrm{A}}_{yx}$ and $\alpha^{\mathrm{A}}_{xy}=-\alpha^{\mathrm{A}}_{yx}$ can be finite and the other components vanish.      It is instructive to rewrite ANC in Eq.~(\ref{Eq:ANC}) by the generalized Mott formula as follows~\cite{Xiao_PhysRevLett.97.026603}, 
\begin{align}
\alpha^{\mathrm{A}}_{ab}  \left(\mu,T \right)&=\frac{1 }{eT}  \int \!\! d\varepsilon  \left(\varepsilon-\mu \right) \frac{\partial f}{\partial \varepsilon}\sigma^{\mathrm{A}}_{ab} \left(\varepsilon,T=0 \right)  \label{Eq:Mott1}\\
&=-\frac{k_\mathrm{B} }{e}  \int \!\! d\varepsilon  s\left(\varepsilon\right) \frac{\partial \sigma^{\mathrm{A}}_{ab}(\varepsilon,T=0)}{\partial \varepsilon}.  \label{Eq:Mott2}
\end{align}
 The well-known Mott relation is obtained as $\alpha^{\mathrm{A}}_{ab} \sim -\frac{\pi^2 k^2_{\mathrm{B}}T}{3e} \frac{\partial \sigma_{ab}(\mu,T=0)}{\partial \mu}$, assuming that $\sigma^{\mathrm{A}}_{ab}(\varepsilon,T=0)\propto \varepsilon-\mu $  for $\varepsilon\sim \mu$ at  low temperature limit.
     
     Figures~\ref{Fig:dos}(b) and (c) display the $\mu$ dependence of AHC and ANC. 
 A shift of the chemical potential corresponds to the rigid band picture, where $\mu>0$ and $\mu<0$ represent electron and hole dopings, respectively.
  Note that agreement of the ANCs calculated via Eqs.~(\ref{Eq:ANC}) and (\ref{Eq:Mott1}) is confirmed numerically as shown in Fig.~\ref{Fig:dos}(c). 
    The calculated values of AHC and ANC are consistent  with previous experimental and theoretical studies~\cite{Liu_shandite_2018,Wang_shandite_2018,Guin_AdvMat.31.1806622,Ozawa_shandite_2019,Ghimire_PhysRevResearch.1.032044,Minami_PhysRevB.102.205128}. 
    Slight differences with previous theoretical calculations may come from   the atomic position of S atoms which is optimized in refs.~\cite{Liu_shandite_2018,Wang_shandite_2018,Guin_AdvMat.31.1806622} while is adopted from the experimental value in the present study. 
 
 Focusing on the ANC for $\mu \le 0$, one can see that the absolute value of the ANC is large at $\mu \sim 0$ and \SI{-0.113}{\electronvolt} with opposite signs. 
   This implies that In-doping induces the sign change and/or the enhancement of the ANE  as will be explicitly shown later.  
   We also note that for qualitative understanding of the $\mu$ dependence of the ANC, the generalized Mott formula in Eq.~(\ref{Eq:Mott2}) is useful. 
 Since the entropy density $s(\varepsilon)$ is an even function with respect to $\varepsilon-\mu$ and rapidly decreases for $|\varepsilon-\mu| \gtrsim k_{\mathrm{B}}T$, 
 one can roughly estimate the sign and magnitude of the ANC from $ \frac{\partial \sigma_{ab}(\varepsilon,T=0)}{\partial \varepsilon}$ for $|\varepsilon-\mu| \lesssim k_{\mathrm{B}}T$.
 On one hand, around $\mu=\SI{0}{\electronvolt}$, $ \sigma_{ab}(\mu,T=0)$ shows steep increase as shown in Fig.~\ref{Fig:dos}(b) and the resulting ANC $\alpha^{\mathrm{A}}_{xy}=\SI{-1.34}{\ampere\per\kelvin\per\metre}$. 
 On the other hand, around  $\mu=\SI{-0.113}{\electronvolt}$, $ \sigma_{ab}(\mu,T=0)$ shows steep decrease and the resulting  ANC $\alpha^{\mathrm{A}}_{xy}=\SI{2.52}{\ampere\per\kelvin\per\metre}$

Next, let us move on to the In-doping dependence. 
 The doping evolutions of the band structures and  net magnetization are shown in  Fig.~\ref{Fig:band}.  
 As mentioned before, in the non-doped case, the half-metallic state is realized and the DOS of the majority spin is metallic, while that of the minority spin has a gap. 
 Hence, when Sn atoms are  substituted by In atoms,  most of  holes are doped into majority spin states.  
  The resulting magnetization decreases almost linearly with respect to the In content $x$ and the system becomes paramagnetic insulator at $x=1.0$ as shown in Fig.~\ref{Fig:band}(h). These behaviors are qualitatively consistent with experimental results~\cite{Kassem_JPSJ2016,Corps_shandite_2015} with slight difference of In-content at which paramagnetic insulating phase emerges.
 Correspondingly, the exchange splitting due to the magnetic ordering decreases with increasing $x$ and 
  the spin splitting of the band structure vanishes for $x=1.0$ as shown in Figs.~\ref{Fig:band}(a)-(f). 
 The half-metallicity is retained in the whole investigated doping range, $x<1.0$.
 
 Results for the transport properties in Co$_3$In$_x$Sn$_{2-x}$S$_2$ are summarized in Fig.~\ref{Fig:AHE}. 
  Figures~\ref{Fig:AHE}(a) and (b)  show the doping dependence of the AHC and ANC. 
For slight hole doped region, the AHC increases  with increasing $x$ and reaches maximum at $x\sim 0.1$. With further increasing $x$, 
the AHC decreases and  vanishes at $x=1.0$ where the system becomes paramagnetic (see also Fig.~\ref{Fig:band}).  
  The AHC sensitively depends on the electronic structure such as the details of the Fermi surfaces and the distribution of the Berry curvature in $\bm{k}$-space.   As a result,  
  the doping dependence of the AHC shows more complicated behavior than that of the magnetic moment~\cite{,Suzuki_PhysRevB.95.094406,Naka_PhysRevB.102.075112}, which linearly decreases with respect to the doping concentration.
 As shown in Fig.~\ref{Fig:AHE}(b), regarding the thermoelectric transport,  the rigid band picture works well for the small doping region [see also Fig.~\ref{Fig:dos}(c)]. 
 The ANC is negative at $x=0$ and changes its sign into positive for the small doping content $x$. The absolute values of $\alpha_{xy}^{\mathrm{A}}/T$  at low temperatures are large at $x\sim 0$ and $0.2$ with values $|\alpha_{xy}^{\mathrm{A}}|/T\sim \SI{0.02}{\ampere.\kelvin^{-2}.\metre^{-1}}$. 
 The ANC  decreases with further increasing $x$, reflecting the suppression of the magnetization.
  To elucidate the enhancement of the ANC for these doping concentrations, the temperature dependence of the ANC is shown in Figs.~\ref{Fig:AHE}(c) and (d). 
 For large doping $x\ge 0.6$, $\alpha^{\mathrm{A}}_{xy}$   shows linear $T$ dependence in the wide temperature range $k_{\mathrm{B}}T\lesssim \SI{0.01}{\electronvolt}$, 
  while for $x=0$ ($0.2$), that exhibits logarithmic $T$ dependence for $ \SI{0.005}{\electronvolt}\lesssim k_{\mathrm{B}}T\lesssim \SI{0.04}{\electronvolt}$ ($ \SI{0.02}{\electronvolt}\lesssim k_{\mathrm{B}}T\lesssim \SI{0.05}{\electronvolt}$). 
  These behaviors could be due to  the stationary points in the nodal rings as demonstrated by Minami \textit{et al.} in ref.~\cite{Minami_PhysRevB.102.205128}. 

  In order to get clear insight into the low temperature behavior of the ANC, let us perform the detailed analysis of the topological properties of the electronic structures which are reflected in the Berry curvature $\Omega_{n,ab}\left(\bm{k}\right)$. 
  As mentioned before, the ANC is dominated by $ \frac{\partial \sigma_{ab}(\varepsilon,T=0)}{\partial \varepsilon}$ for  $|\varepsilon-\mu| \lesssim k_{\mathrm{B}}T$, whose explicit form is given as follows~\cite{Ghimire_PhysRevResearch.1.032044},  
\begin{align}
&\frac{\partial \sigma^{\mathrm{A}}_{ab}\left(\varepsilon, T=0 \right)}{\partial \varepsilon } = -\frac{e^2}{\hbar}\sum_{n}\int_{\varepsilon_{n\bm{k}}
=\varepsilon} \!\! \frac{dS_{\bm{k}}}{\left(2\pi \right)^3} \frac{\Omega_{n,ab} \left(\bm{k}\right)}{\left|\bm{\nabla}_{\bm{k}}\varepsilon_{n\bm{k}}\right|}, \label{Eq:dersigma}
\end{align}
where $\int_{\varepsilon_{n\bm{k}}=\varepsilon} \!\! dS_{\bm{k}}$ represents the $\bm{k}$-integration over the isoenergy surfaces for $\varepsilon_{n\bm{k}}=\varepsilon$. 
  From Eqs.~(\ref{Eq:Mott2}) and (\ref{Eq:dersigma}),  one can see that the qualitative behavior of the ANC at low temperatures are governed by 
 the distribution of the  $ \frac{\Omega_{n,ab} \left(\bm{k}\right)}{\left|\bm{\nabla}_{\bm{k}}\varepsilon_{n\bm{k}}\right|}$  near the Fermi surfaces.  
 In Fig.~\ref{Fig:FS_Berry}(a), we show the Fermi surfaces with $ \frac{\Omega_{n,ab} \left(\bm{k}\right)}{\left|\bm{\nabla}_{\bm{k}}\varepsilon_{n\bm{k}}\right|}$. 
   The value of $ \frac{\Omega_{n,ab} \left(\bm{k}\right)}{\left|\bm{\nabla}_{\bm{k}}\varepsilon_{n\bm{k}}\right|}$ is positive on the small portions of the Fermi surfaces for $x=0$,  
  while on  more wide region, $ \frac{\Omega_{n,ab}\left(\bm{k}\right)}{\left|\bm{\nabla}_{\bm{k}}\varepsilon_{n\bm{k}}\right|}$ becomes negative.  
   As a result, the $\bm{k}$-integration over the Fermi surfaces of  $\frac{\Omega_{n,ab}\left(\bm{k}\right)}{\left|\bm{\nabla}_{\bm{k}}\varepsilon_{n\bm{k}}\right|}$ is negative, resulting in   $\alpha^{\mathrm{A}}_{xy}<0$ for Co$_3$Sn$_2$S$_2$ as shown in Fig.~\ref{Fig:AHE}(b). 
  On the other hand, for $x=0.2$, the area of the Fermi surfaces on which 
  $\frac{\Omega_{n,ab} \left(\bm{k}\right)}{\left|\bm{\nabla}_{\bm{k}}\varepsilon_{n\bm{k}}\right|}$ is negative becomes small
   and the resulting ANC is positive.  
     The absolute value of $ \frac{\Omega_{n,ab} \left(\bm{k}\right)}{\left|\bm{\nabla}_{\bm{k}}\varepsilon_{n\bm{k}}\right|}$ decreases with further increasing $x$, and the ANCs  for large $x$ have small values [see also Fig.~\ref{Fig:AHE}(b)].  

We here demonstrate that the intensity of the Berry curvature on Fermi surfaces is closely related to the nodal lines.  As shown in Fig.~\ref{Fig:FS_Berry}(b), the nodal lines appear on the mirror symmetry planes in the absence of the spin-orbit coupling. The spin-orbit coupling gaps out the nodal lines but leaves the energy $\sim \SI{0.065}{\electronvolt}$, resulting in the Weyl nodes for $x=0$~\cite{Liu_shandite_2018,Wang_shandite_2018}. In this case, the energy bands characterized by different eigen values $\pm 1$ of the mirror symmetry operator produce  nodal lines due to its crossing  [see Fig.~\ref{Fig:FS_Berry}(c)]. 
 This origin of nodal lines is similar to those in archetypal nonmagnetic nodal line semimetals Ca$_3$P$_2$~\cite{Xie_APLMater_Ca3P2_2015,Chan_PhysRevB.93.205132} and CaAg$X$ ($X$=P, As)~\cite{Yamakage_JPSJ.85.013708}. 
   
 The hole doping decreases the Fermi energy and, as a result, shift the Weyl nodes away from the Fermi level.  The nodal rings are located around the Fermi energy for small hole doping and  
 for $x\lesssim 0.2$, the nodal lines surrounding the L-point cross the Fermi surfaces as shown in Fig.~\ref{Fig:FS_Berry}(b). The band at L-point near the Fermi energy with mirror eigen value $+1$ shifts upward in energy with increasing $x$ and cross with the band having mirror eigen value $-1$ for $x\sim 0.3$, as shown in Fig.~\ref{Fig:FS_Berry}(c). Then, the nodal lines are split into two rings, as shown in Fig.~\ref{Fig:FS_Berry}(b), with the annihilation of Weyl nodes in the presence of the spin-orbit coupling. 
 The nodal lines still cross the Fermi surfaces for $x< 0.4$ and give the significant contribution to the ANC. 
 For $x\gtrsim 0.6$, the nodal lines are located far from the Fermi level, resulting in the small Berry curvature on the whole Fermi surfaces. 
One can clearly see that the intensity of the Berry curvature is large near the nodal lines.

\section{Summary and discussions}

 In the present paper, we  investigated the  magnetic and transport properties in Co$_3$In$_x$Sn$_{2-x}$S$_2$ based on first-principles calculations in which   In-doping effect is treated within the virtual crystal approximation. 
 We show that the anomalous Hall and Nernst conductivities show complicated behaviors with varying $x$ against linearly reduced magnetization with respect to  In content $x$ due to the half-metallic electronic states.
  The Nernst conductivity has large values for $x\sim 0$ and $0.2$ with opposite signs and show logarithmic temperature dependence consistently with the previous theoretical study~\cite{Minami_PhysRevB.102.205128}. 
 We also clarify that the low temperature behavior of the anomalous Nernst conductivity can be understood by the distribution of the Berry curvature  divided by the Fermi velocity. 
  The close relationships between the Berry curvature, nodal rings, and anomalous Nernst effect are explicitly demonstrated.  
   The intensity of the Berry curvature has a large value near the nodal rings for small hole doping, leading to a large  anomalous Nernst conductivity.  The In-doping induces the reconstruction of the nodal rings and moves the nodal lines far away from the Fermi level, 
    resulting in the small Berry curvature on the Fermi surfaces. 
 Our results  give a qualitative understanding of the thermoelectric transport in Co$_3$In$_x$Sn$_{2-x}$S$_2$ and encourage  experimental measurements of the anomalous Nernst effect in Co$_3$In$_x$Sn$_{2-x}$S$_2$. 
 
 Lastly, we comment on the important issues which are not addressed in this study. 
 In the present paper,  we have  focused on the intrinsic contribution to the anomalous transports and  neglected the extrinsic effects such as  side-jump and skew-scattering~\cite{Nagaosa_RevModPhys.82.1539}. 
 Effects of the structural disorder, however,  is inevitable in doped compounds in general and  might give considerably large extrinsic contribution to the thermoelectric transport pointed by 
   several authors~\cite{Jianlei_adfm.202000830,Ding_PhysRevX.9.041061,Papaj_2020arXiv200807974P}. 
  A quantitative study on extrinsic contributions using  first-principles calculations is a remaining issue in a future work.

\begin{acknowledgments}
This research was supported by JSPS KAKENHI Grants Numbers 
JP15H05883 (J-Physics), JP18H04230, JP19H01842, JP20H01830, JP20H05262, JP20K05299, and JP20K21067. 
We also acknowledge  supports from JST PRESTO Grant number JPMJPR17N8 and JST CREST Grant number JPMJCR18T2.
A part of the numerical calculations was carried out using MASAMUNE-IMR of the Center for Computational Materials Science, Institute for Materials  Research, Tohoku University.
YY thanks H. Kusunose for sharing computer facilities. 
\end{acknowledgments}

\bibliography{ref}% Produces the bibliography via BibTeX.

\end{document}